\documentclass[12pt]{iopart}
\usepackage{color}
\usepackage{graphicx}
\usepackage{sidecap}
\usepackage{dcolumn}
\usepackage{subfigure}
\usepackage{epsfig}
\usepackage[all]{xy}
\def\cao{\c c\~ao}

\begin{document}

\title[A Galerkin-Collocation domain decomposition method]{A Galerkin-Collocation domain decomposition method: application to the evolution of cylindrical gravitational waves}
\author{W O Barreto$^{1, 2}$, J A Crespo$^2$, H P de Oliveira$^2$ and E L Rodrigues$^3$}
\address{$^1$ Centro de F\'{\i}sica Fundamental, Universidad de Los Andes, M\'erida 5101,  Venezuela}
\address{$^2$ Departamento de F\'{\i}sica Te\'orica - Instituto de F\'{\i}sica
A. D. Tavares, Universidade do Estado do Rio de Janeiro, 
R. S\~ao Francisco Xavier, 524. Rio de Janeiro, RJ, 20550-013, Brazil}
\address{$^3$ Instituto de Bioci\^encias - Departamento de F\'isica, 
Universidade Federal do Estado do Rio de Janeiro, 
Av. Pasteur, 458 - Urca. Rio de Janeiro, RJ, 22290-040, Brazil}
\date{\today}

\begin{abstract}
We present a Galerkin-Collocation domain decomposition algorithm applied to the evolution of cylindrical unpolarized gravitational waves. We show the effectiveness of the algorithm in reproducing initial data with high localized gradients and in providing highly accurate dynamics. We characterize the gravitational radiation with the standard Newman-Penrose Weyl scalar $\Psi_4$. We generate wave templates for both polarization modes, $\times$ and $+$, outgoing and ingoing, to show how they exchange energy nonlinearly. In particular, considering an initially ingoing $\times$ wave, we were able to trace a possible imprint of the gravitational analog of the Faraday effect in the generated templates.
\end{abstract}
\submitto{\CQG}
\maketitle

\section{Introduction}%

In the last years, we have witnessed the growing popularity of spectral methods in numerical relativity with applications in a large variety of problems \cite{grandc_novak}. The main advantage of spectral methods when compared with the traditional finite difference methods is the superior accuracy for a fixed number of grid points \cite{kidder}. In particular, for smooth functions, the convergence rate exhibited by spectral methods is exponential.  On the other hand, the accuracy of spectral methods in solving partial differential equations is drastically reduced in the case the solutions have localized regions of rapid variations, or if the spatial domain has a complex geometry \cite{canuto,gottlieb_hesthaven}. 

Multidomain techniques \cite{canuto,gottlieb_orszag,kopriva_86,orszag_80,kopriva_89} or simply the domain decomposition method is a beautiful and efficient strategy to improve the accuracy of spectral approximations for the cases mentioned above. The spatial domain is divided into two or more subdomains, where we can establish spectral approximations of a function in each subdomain together with the matching or transmissions conditions across the subdomains boundaries. In this work, we implemented a version of the spectral
fixed mesh refinement method.

In numerical relativity, the first applications of the domain decomposition technique involved the determination of the stationary configurations \cite{gourgoulhon} and the initial data problem \cite{pfeiffer,ansorg}, mainly for binaries of black holes. For the time-dependent systems, the spectral domain decomposition was implemented within the SpEC \cite{spec} and LORENE \cite{lorene} codes to deal with the gravitational collapse, the dynamics of stars and the evolution of single \cite{kidder_scheel_teuk} and binary black holes \cite{szilagyi_lindblom_scheel}. A more detailed approach for the spectral multi-domain codes is found in Refs. \cite{hemberger} and in the SXS collaboration \cite{sxs}.

The crucial step for the domain decomposition technique is the treatment of the transmission conditions between subdomains. In the case of non-overlapping subdomains and time-independent situations,  as the elliptic initial data problem, the smoothness of any function and its normal derivative is guaranteed on the interface of the subdomains. In the case of hyperbolic problems, in general, are required other interface conditions \cite{kopriva_86,canuto,patera} to preserve stability.

The aim of the present work is twofold. First, we present an innovative Galerkin-Collocation domain decomposition algorithm to evolve general cylindrical gravitational waves. We have divided the physical spatial domain into two subdomains and introduced the corresponding computational subdomains which are the loci of the collocation points. The communication between the physical and the computational subdomains is established by distinct mappings that cover the whole spatial domain. Second, we explore the consequences of the interaction between the gravitational wave polarization modes by generating the wave templates associated with the polarization modes at the radiation zone. Although cylindrical gravitational waves do not represent a real physical situation, they provide a useful theoretical laboratory to investigate the interaction of the polarization wave modes \cite{piran}.

In the context of cylindrical symmetry, exact solutions are representing polarized, and unpolarized waves in the form of Weber-Wheeler \cite{weber_wheeler} and Xanthopoulos \cite{xanthopoulos} solutions are known. However, the present numerical strategy can be useful for studying more general spacetimes admitting gravitational waves with no available analytical solutions.

We organized the paper as follows. In  Section 2, we present the basic equations of the general cylindrical gravitational waves. The Galerkin-Collocation domain decomposition method is described in details in Section 3, along with the numerical scheme to evolve the field equations. 
Section 4 deals with the validation of the code through numerical tests of convergence. We have discussed the physical aspects in Section 5 by establishing the version of the peeling theorem for cylindrical spacetimes. In the sequence, we obtain the expression for the Weyl scalar, $\Psi_4$, that determines the templates of the gravitational waves at the wave zone, with a special interest in those resulting from the interaction between both polarization modes. We close this work with some final remarks in Section 6.

\section{Basic equations}%

We consider the general cylindrical line element proposed by Kompaneets \cite{kompaneets} and Jordan {\it et al.} \cite{jordan}. The metric is written initially in the  $3+1$ formulation, but we adopt here the version using null coordinates,

\begin{eqnarray}
ds^2 =&& -{\rm e}^{2(\gamma - \psi)} (du^2 + 2~du~d\rho) + {\rm e}^{2\psi}(dz+\omega d\phi )^{2} + \rho^{2}{\rm e}^{-2\psi}d\phi ^{2}, \label{eq1}
\end{eqnarray}

\noindent where $u$ is the retarded null coordinate that foliates the spacetime in hypersurfaces $u=\mathrm{constant}$ and $(\rho,z,\phi)$ are the usual cylindrical coordinates. The metric functions $\psi$, $\omega$, $\gamma$ depend on $u$ and $\rho$. As a well-known important aspect of cylindrical spacetimes \cite{thorne}, the functions $\psi$ and $\omega$ represent the two dynamical degrees of freedom of the gravitational field, in which $\psi$ accounts for the polarization mode $+$ while $\omega$ the polarization mode $\times$ \cite{thorne}. The function $\gamma$ plays the role of the gravitational energy of the system and gives the total energy per unit length enclosed within a cylinder of radius $\rho$ at the time $u$. The function $\gamma$ is related to the $C$-energy \cite{thorne,stachel,goncalves}, which satisfies the conservation law $P^i_{;i}=0$, where $P^i$  is the C-energy flux vector \cite{thorne}. It is important to mention that the Einstein-Rosen gravitational waves propagate the $C$ energy which has reinforced the radiative character of this solution.

Following Refs. \cite{clarke_dinverno,celestino} it is convenient to introduce a new radial coordinate $y$ by
\begin{eqnarray}
\rho=y^2, \label{eq2} 
\end{eqnarray}

\noindent and to define the new fields $\bar{\psi}$ and $\bar{\omega}$, respectively by 
\begin{eqnarray}
\bar{\psi}=y \psi \label{eq3}, \\
\nonumber \\
\bar{\omega}=\frac{\omega}{y}. \label{eq4}
\end{eqnarray}

\noindent Thus, the field equations in terms of the new fields $\bar{\psi}$ and $\bar{\omega}$ are expressed by
{\small{
\begin{eqnarray}
&& y\bar{\psi}_{,uy} - \frac{{\rm e}^{{4\bar{\psi}}/{y}}}{2y} (\bar{y\omega})_{,y} \bar{\omega}_{,u} - \frac{1}{4}\left[y\left(\frac{\bar{\psi}}{y}\right)_{,y}\right]_{,y} + 
\frac{{\rm e}^{{4\bar{\psi}}/{y}}}{8y^3} (y\bar{\omega})_{,y}^2 = 0, \label{eq5} \\
&& y\bar{\omega}_{,uy} + \frac{2}{y} (y\bar{\omega})_{,y}\bar{\psi}_{,u} + 2 y \left(\frac{\bar{\psi}}{y}\right)_{,y} \bar{\omega}_{,u} - \frac{y^2}{4}\left[\frac{(y\bar{\omega})_{,y}}{y^3}\right]_{,y} - 
\frac{1}{y}(y\bar{\omega})_{,y}\left(\frac{\bar{\psi}}{y}\right)_{,y} = 0, \label{eq6}
\end{eqnarray}}}

\noindent obtained from the components $R_{zz}=0$ and $R_{\phi\phi}=0$. Here the subscripts $u$ and $y$ denote partial derivatives with respect to these coordinates.

The dynamics of cylindrical spacetimes is fully described by the coupled wave equations (\ref{eq5}) and (\ref{eq6}) for the gravitational potencials $\bar{\psi},\bar{\omega}$ starting with the initial data functions $\bar{\psi}_0(y)=\bar{\psi}(u_0,y)$ and $\bar{\omega}_0(y)=\bar{\omega}(u_0,y)$. These initial distributions are free of any constraint according with the characteristic scheme we adopt here.

The metric function $\gamma$ satisfies the remaining field equations $R_{yy}=0$ and $R_{uu}-R_{uy}=0$, or
\begin{eqnarray}
\gamma_{,y} &=& \frac{y}{2}\left(\frac{\bar{\psi}}{y}\right)_{,y}^2 + \frac{\mathrm{e}^{{4\bar{\psi}}/{y}}}{8y^3}\left(y\bar{\omega}\right)^2_{,y}, \label{eq7}\\
\nonumber \\
\gamma_{,u} &=& \bar{\psi}_{,u}\left(\frac{\bar{\psi}}{y}\right)_{,y} - 2 \bar{\psi}_{,u}^2 + \frac{\mathrm{e}^{{4\bar{\psi}}/{y}}}{4y^2}[\bar{\omega}_{,u}(y\bar{\omega})_{,y}
-2y^2\bar{\omega}_{,u}^2].  \label{eq8}
\end{eqnarray}

\noindent Thus, the evolution of $\gamma(u,y)$ is determined after solving the wave equations (\ref{eq5}) and (\ref{eq6}). 

In general, we establish the conditions to guarantee the well-behaved coordinates as well as the regularity of the spacetime. After a careful inspection of the field equations (\ref{eq5}) and (\ref{eq6}), the conditions of regularity and flatness of the metric near the origin $y=0$ impose that
\begin{eqnarray}
&&\bar{\psi}(u,y)=\mathcal{O}(y),  \label{eq9}\\ 
\nonumber \\
&&\bar{\omega}(u,y)=\mathcal{O}(y^3).  \label{eq10}
\end{eqnarray}

\noindent The other conditions are specified at the future null infinity, $\mathcal{J}^+$ ($y=\infty$). The asymptotic analysis \cite{stachel} of the wave equations (5) and (6) leads to
\begin{eqnarray}
&&\bar{\psi}(u,y) = \bar{\psi}_{\infty}(u) + \mathcal{O}(y^{-1}), \label{eq11}\\ 
\nonumber \\
&&\bar{\omega}(u,y)=\bar{\omega}_{\infty}(u) + \mathcal{O}(y^{-2}),\label{eq12}
\end{eqnarray}

\noindent where $\bar{\psi}_{\infty}(u)$ and $\bar{\omega}_{\infty}(u)$ are arbitrary functions{, because the spacetime is not asymptotically flat, which is intrinsic to cylindrical symmetry}. 

{Before closing this section, we want to stress the following  asymmetry in the wave equations (\ref{eq5}) and (\ref{eq6}), that will be useful for understanding the numerical results. If initially we have a pure wave mode $\bar{\psi}$, it follows that Eq. (\ref{eq6}) is identically satisfied.
 Thus, a pure initial mode $\bar\psi$ does not excite $\bar\omega$ (observe that the resulting Eq. (\ref{eq5}) becomes linear under these conditions). However, a pure initial mode $\bar\omega$ does not vanish Eq. (\ref{eq5}). As a consequence, this  pure mode $\bar\omega$ will excite a non-trivial $\bar\psi$ producing mode mixing in the evolution.}

\begin{figure}[h]
\begin{center}
\includegraphics*[width=8cm,height=6cm]{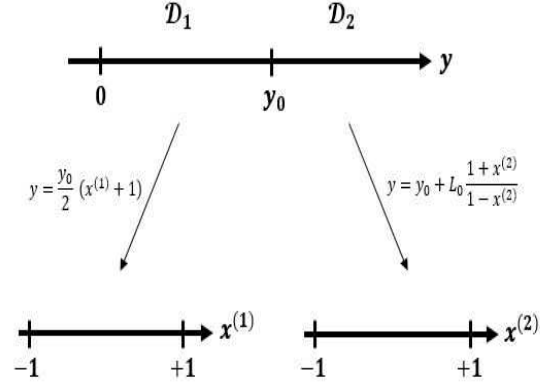}
\end{center}
{\renewcommand{\baselinestretch}{1}
\vspace{-0.5cm}
\caption{Basic scheme showing the subdomains $\mathcal{D}_1: 0 \leq y \leq y_0$ and $\mathcal{D}_2: y_0 \leq y < \infty$ and the corresponding maps to the computational variable $x^{(A)}$ where $-1 \leq x^{(A)} \leq 1$ with $A=1,2$.}}
\end{figure}

\section{The Galerkin-Collocation domain decomposition method}%

We present here the domain decomposition algorithm for the dynamics of general cylindrical gravitational waves. In Ref. \cite{celestino} we have implemented a single domain method for the same problem. However, we have made a comment about the lack of exponential convergence using the single domain code to testing it against the analytical vacuum Xanthopoulos solution \cite{xanthopoulos} that contains both gravitational degrees of freedom $\bar{\psi}$ and $\bar{\omega}$ describing gravitational waves with polarization modes $+$ and $\times$. In general, the exact profiles exhibit rapid variations of the corresponding fields that spoil the exponential convergence for the case of one domain code. A domain decomposition algorithm improves the convergence as we indicated briefly in \cite{celestino}. In what follows, we present the details.

We begin by dividing the spatial domain in two subdomains: $\mathcal{D}_1: 0 \leq y \leq y_0$ and $\mathcal{D}_2: y_0 \leq y < \infty$, where $y=y_0$ is the interface between both domains. The equivalent computational subdomains are indicated by $\mathcal{D}_{A}: -1 \leq x^{(A)} \leq 1$, with $A=1,2$ (cf. Fig. 1). We have chosen the following maps that connect the computational and physical subdomains:
\begin{eqnarray}
\mathcal{D}_1:&& \; y =\frac{y_0}{2}(1+x^{(1)}),\label{eq13}\\
\nonumber \\
\mathcal{D}_2:&& \; y =y_0+L_0\frac{(1+x^{(2)})}{(1-x^{(2)})},\label{eq14}
\end{eqnarray} 

\noindent where $L_0$ is the map parameter. 

In each subdomain $\mathcal{D}_{A}$, $A=1,2$, we establish the spectral approximations for the gravitational potentials $\bar{\psi}$ and $\bar{\omega}$ as
\begin{eqnarray}
&& \bar{\psi}^{(A)}(u,y) = \sum_{k=0}^{N^{(A)}_\psi}\,a^{(A)}_k(u) \Psi^{(A)}_k(y), \label{eq15} \\
\nonumber \\
&& \bar{\omega}^{(A)}(u,y) = \sum_{k=0}^{N^{(A)}_\omega}\,b^{(A)}_k(u) \Phi^{(A)}_k(y), \label{eq16}
\end{eqnarray}

\noindent where $N^{(A)}_\psi$ and $N^{(A)}_\omega$ are the truncations orders, not necessarily equal, that dictate the number of unknow modes $a^{(A)}_j(u)$ and $b^{(A)}_k(u)$, respectively. According to the Galerkin method, the basis functions ${\Psi^{(A)}_j(y)}$ and ${\Phi^{(A)}_k(y)}$ satisfy the conditions (\ref{eq9}) - (\ref{eq12}) by combining conveniently the rational Chebyshev polynomials defined in each subdomain. The rational Chebyshev polynomials \cite{boyd} defined in each subdomain are
\begin{eqnarray}
TL^{(1)}_k(y) &=& T_k\left(x^{(1)}=\frac{2y}{y_0}-1\right), \label{eq17}\\
\nonumber \\
TL^{(2)}_k(y) &=& T_k\left(x^{(2)}=\frac{y-y_0-L_0}{y-y_0+L_0}\right). \label{eq18}
\end{eqnarray}

\noindent where $T_k(x)$ represents the standard Chebyshev polynomial of $kth$-order. We present below the basis functions:
{\small
\begin{eqnarray}
\Psi^{(1)}_k(y) &=& \frac{1}{2}(TL^{(1)}_{k+1}(y)+TL^{(1)}_k(y)),\; \nonumber \\
\Psi^{(2)}_k(y) &=& TL^{(2)}_{k}(y), \nonumber \\
\Phi^{(2)}_k(y) &=& \Psi^{(2)}_k(y), 
\end{eqnarray}}

\noindent and the corresponding expression to $\Phi^{(1)}_k(y)$ can be found in  Appendix A.

We have followed the domain decomposition method straightforwardly for hyperbolic problems according to Gottlieb and Orszag \cite{gottlieb_orszag}. In their approach, the junction or transmission conditions are 
{\small{
\begin{eqnarray}
\bar{\psi}^{(1)}(u,y_0) &=&\bar{\psi}^{(2)}(u,y_0), \nonumber \\ \nonumber \\
\left(\frac{\partial \bar{\psi}^{(1)}}{\partial y}\right)_{y_0} &=& \left(\frac{\partial \bar{\psi}^{(2)}}{\partial y}\right)_{y_0}, \nonumber \\
\nonumber\\
\bar{\omega}^{(1)}(u,y_0) &=&\bar{\omega}^{(2)}(u,y_0), \nonumber \\ \nonumber \\
\left(\frac{\partial \bar{\omega}^{(1)}}{\partial y}\right)_{y_0} &=& \left(\frac{\partial \bar{\omega}^{(2)}}{\partial y}\right)_{y_0}.
\label{eq19}
\end{eqnarray}}}

\noindent Taking into account the spectral approximations of the metric functions (\ref{eq15}) and (\ref{eq16}) into the above transmission conditions, we obtain four linear equations involving the coefficients $a^{(A)}_k(u)$ and $b^{(A)}_k(u)$, $A=1,2$. Furthermore, these relations are used to reduce the total number of independent coefficients of $\bar{\psi}^{(A)}$ and $\bar{\omega}^{(A)}$ to $N^{(1)}_\psi+N^{(2)}_\psi+2-2=N^{(1)}_\psi+N^{(2)}_\psi$ and $N^{(1)}_\omega+N^{(2)}_\omega+2-2=N^{(1)}_\omega+N^{(2)}_\omega$, respectively.

We proceed by establishing the residual equations by substituting the approximations (\ref{eq15}) and (\ref{eq16}) into the field equations (\ref{eq5}) and (\ref{eq6}) which yields
\begin{eqnarray}
\mathrm{Res}_{\psi}^{(A)}(u,y)&=& y\bar{\psi}^{(A)}_{,uy} - \frac{{\rm e}^{{4\bar{\psi}^{(A)}}/{y}}}{2y} (\bar{y\omega}^{(A)})_{,y} \bar{\omega}^{(A)}_{,u} - \frac{1}{4}\left[y\left(\frac{\bar{\psi}^{(A)}}{y}\right)_{,y}\right]_{,y} \nonumber\\
&&+ \frac{{\rm e}^{{4\bar{\psi}^{(A)}}/{y}}}{8y^3} (y\bar{\omega}^{(A)})_{,y}^2, \label{eq20} \\
\nonumber \\
\mathrm{Res}^{(A)}_{\omega}(u,y)&=&y\bar{\omega}^{(A)}_{,uy} + \frac{2}{y} (y\bar{\omega}^{(A)})_{,y}\bar{\psi}^{(A)}_{,u} + 2 y \left(\frac{\bar{\psi}^{(A)}}{y}\right)_{,y} \bar{\omega}^{(A)}_{,u} - \frac{y^2}{4}\left[\frac{(y\bar{\omega}^{(A)})_{,y}}{y^3}\right]_{,y} \nonumber \\
&&-\frac{1}{y}(y\bar{\omega}^{(A)})_{,y}\left(\frac{\bar{\psi}^{(A)}}{y}\right)_{,y}. 
\end{eqnarray}

\noindent In general the residuals $\mathrm{Res}^{(A)}_\psi(u,y)$ and $\mathrm{Res}^{(A)}_{\omega}(u,y)$ do not vanish since $\bar{\psi}^{(A)}$ and $\bar{\omega}^{(A)}$ are approximations to the exact corresponding gravitational potentials. In accordance with the numerical strategy we are adopting, we use the Collocation method in the sense that the residual equations vanish at the collocation or grid points. Schematically we may write
\begin{eqnarray}
\mathrm{Res}^{(A)}_{\psi}(u,y_k)&=& 0,\;\; k=0,1,..,N^{(A)}_{\psi}-1, \label{eq21}\\
\nonumber \\
\mathrm{Res}^{(A)}_{\omega}(u,y_k)&=& 0,\;\; k=0,1,..N^{(A)}_{\omega}-1, \label{eq22}
\end{eqnarray}

\noindent where $y_k$ denotes the collocation points in the physical subdomains. We have calculated these collocation points from the Chebyshev-Gauss points $x^{(A)}_k$
\begin{equation}
x^{(A)}_k = \cos\left(\frac{(2k+1)\pi}{2N^{(A)}}\right),\;\;k=0,1,..,N^{(A)}-1, \label{eq23}
\end{equation}

\noindent and from the maps (\ref{eq13}) and (\ref{eq14}). Here $N^{(A)}$ denotes either $N^{(A)}_\psi$ or $N^{(A)}_\omega$. 

We have approximated the field equations into a set of ordinary differential equations written in the following matricial form
\begin{eqnarray}
\textbf{M} 
\left(
\begin{array}{c}
\partial \bar{\psi}^{(1)}_k \\ 
\partial \bar{\psi}^{(2)}_k \\ 
\partial \bar{\omega}^{(1)}_j \\
\partial \bar{\omega}^{(2)}_j 
\end{array}
\right ) = \mathbf{B}, \label{eq24}
\end{eqnarray}

\noindent for all $k=0,1,..,N^{(A)}_\psi-1$ and $j=0,1,.,N^{(A)}_\omega-1$. In the above expression we have
{\small
\begin{eqnarray}
\partial \bar{\psi}^{(A)}_k(u) \equiv \left(\frac{\partial \bar{\psi}^{(A)}}{\partial u}\right)_k = \sum_{i=0}^{N_\psi}\,a^{(A)}_{i,u}(u) \Psi^{(A)}_i(y_k), \label{eq25}\\
\nonumber \\
\partial \bar{\omega}^{(A)}_j(u) \equiv \left(\frac{\partial \bar{\omega}^{(A)}}{\partial u}\right)_j = \sum_{i=0}^{N_\psi}\,b^{(A)}_{i,u}(u) \Phi^{(A)}_i(y_k),
 \label{eq26}
\end{eqnarray}}

\noindent where $\partial \bar{\psi}^{(A)}_k(u)$ and $\partial \bar{\omega}^{(A)}_j(u)$ are the values of the derivatives of $\bar{\psi}^{(A)}$ and $\bar{\omega}^{(A)}$ with respect to $u$ at the collocation points. Note that these values are related to the time derivatives of the unknown modes $a^{(A)}_{k,u}(u),b^{(A)}_{j,u}(u)$. The matrices $\textbf{M}$ and $\textbf{B}$ depend on the unknown modes $a^{(A)}_k(u),b^{(A)}_j(u)$ as well the values of $\bar{\psi}^{(A)}$ at the collocation points, or 
{\small
\begin{eqnarray}
\bar{\psi}^{(A)}_k(u) \equiv \bar{\psi}^{(A)}(u,y_k) = \sum_{i=0}^{N_\psi}\,a^{(A)}_i(u) \Psi^{(A)}_i(y_k), \label{eq27}
\end{eqnarray}}

\noindent that provides a set of relations between the values and the unknown modes. The integration is performed as follows: starting from the initial modes $a^{(A)}_k(u_0),b^{(A)}_k(u_0)$ we can determine the initial values $\bar{\psi}^{(A)}_k(u_0)$ as well the initial matrices $\textbf{M},\textbf{B}$. With  the matrices $\textbf{M},\textbf{B}$ evaluated at $u=u_0$, we obtain  the initial values $\partial \bar{\psi}^{(A)}_k(u_0),\partial \bar{\omega}^{(A)}_j(u_0)$ from the dynamical system (26). According to the relations (27) and (28) we can determine $a^{(A)}_{k,u}(u_0),b^{(A)}_{k,u}(u_0)$, and as a consequence, the modes $a^{(A)}_{k},b^{(A)}_{k}$ are calculated at the next step and the whole process repeats providing the evolution of the system. In this case, we have used a fourth-order Runge-Kutta integrator.

%

\section{Numerics}%

We start comparing the domain decomposition algorithm with the single domain one \cite{celestino} by reproducing some exact initial profiles of the gravitational potentials numerically. In the first example, we have considered the exact Weber-Wheeler \cite{weber_wheeler} solution that corresponds to the case $\omega=0$. The exact initial profile $\psi_{\mathrm{exact}}(u_0,y)$ (see the exact solution in Appendix B) has two parameters, $A_0$ and $a$ identified as the amplitude and the width of the wave, respectively. We have set $A_0=1$, $a=1$ and $u_0=-10$ to characterize an initial profile with a steep slope. In Fig. 2, we generated the numerical profiles (circles) using the single domain and the domain decomposition algorithms with $N_\psi=60$ and $N^{(1)}_\psi+N^{(2)}_\psi=40$ collocation points, respectively. It is clear the effectiveness of the domain decomposition in reproducing the initial profile with a smaller number of collocation points.
\begin{figure}[htb]
	\begin{center}
		\subfigure[]{\scalebox{.65}{\input{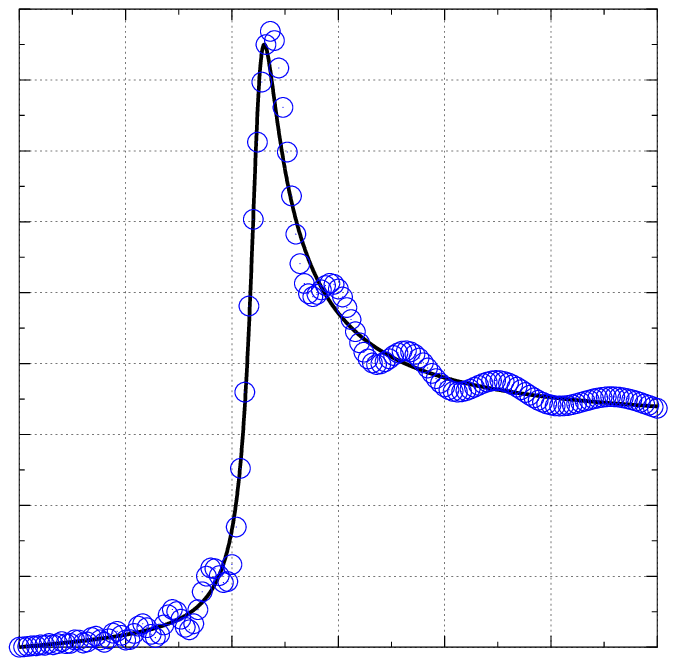}}}
		\hspace{-2cm}
		\subfigure[]{\scalebox{.65}{\input{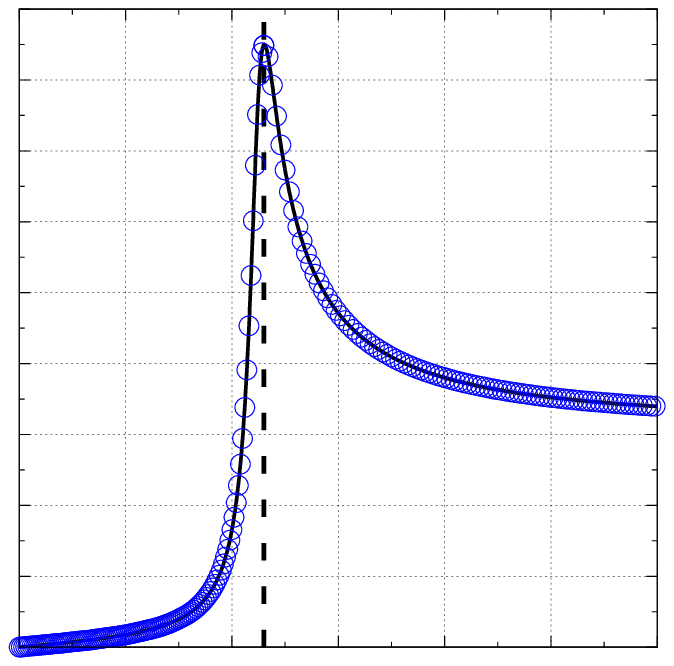}}}
	\end{center}
	\caption{Initial exact (lines) and approximated (circles) profiles of the Weber-Wheeler solution evaluated at $u_0=-10$. We use the single domain in (a) and the domain decomposition in (b). Here $L_0=y_0=2.3$.}
\end{figure}
\begin{figure*}[htb]
\begin{center}
\subfigure[]{\scalebox{0.65}{\input{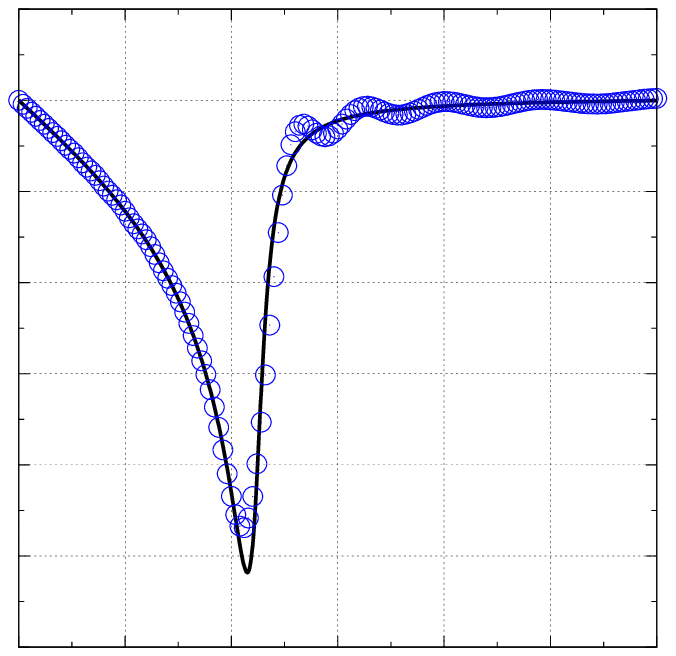}}\hspace{-2.5cm}\scalebox{0.65}{\input{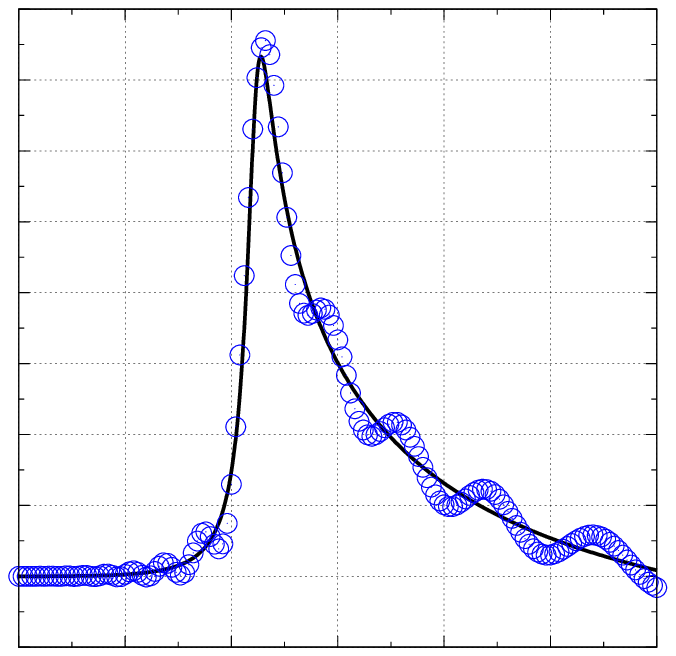}}}
\hspace{-1cm}
\subfigure[]{\scalebox{0.65}{\input{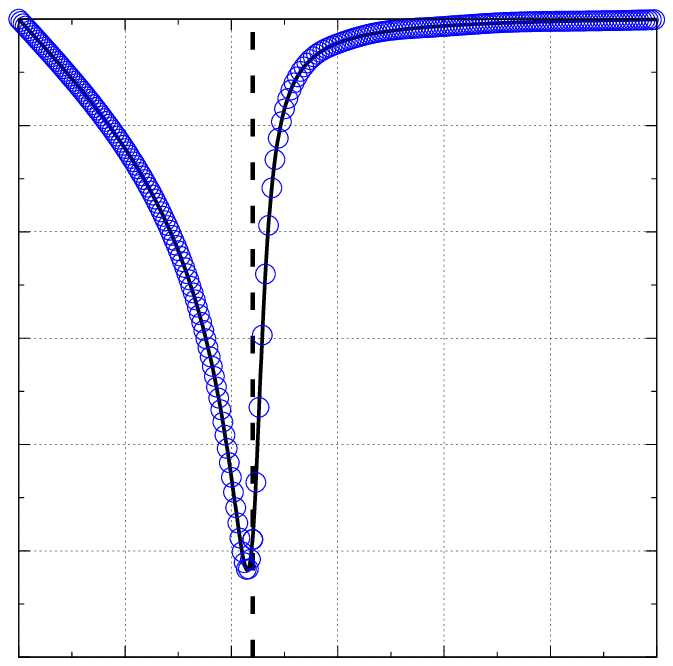}}\hspace{-2.5cm}\scalebox{0.65}{\input{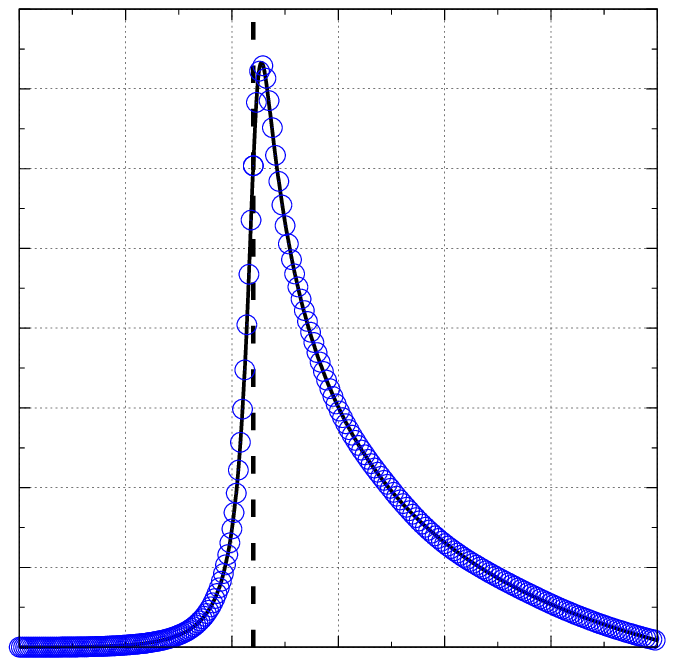}}}
\end{center}
\vspace{-0.7cm}
\caption{Initial exact (lines) and approximated (circles) profiles of the Xanthopoulos solution evaluated at $u_0=-10$. In (a) and (b) we obtained the approximate solutions using the single domain and the decomposition domains algorithms, respectively. Here $y_0=2.2$.}
\end{figure*}

In the second example, we considered the exact profiles of both gravitational potentials from the Xanthopoulos exact solution \cite{xanthopoulos} (Appendix B) evaluated at $u_0=-10$ to produce profiles with steep slopes. Fig. 3 shows qualitatively the efficiency of the domain decomposition scheme (truncation orders $N^{(A)}_\psi=20,N^{(A)}_\omega=20$, $A=1,2$) over the single domain procedure ($N_\psi=N_\omega=60$). We placed the interface at $y_0=2.2$ that approximately is close to the location of the steep slope for the better accuracy in reproducing the initial profiles. 

Using the Bondi's formula presented by Stachel \cite{stachel} we made the basic convergence test for the evolved cylindrical gravitational waves. For the sake of completeness the referred  formula of Bondi is  
\begin{equation}
\frac{d M}{d u}=-\left[\left(\frac{d c_1}{d u}\right)^2+\left(\frac{d c_2}{d u}\right)^2\right],\label{eq28}
\end{equation}

\noindent where $M(u)$ is the Bondi mass aspect (indeed mass per unit of length), ${d c_1}/{du}$ and ${d c_2}/{du}$  are the news functions associated to each degree of freedom of the gravitational waves. These quantities are calculated according to 
\begin{eqnarray}
M(u) &=& \frac{1}{2} \lim_{y \rightarrow \infty}\,\gamma, \label{eq29} \\
\nonumber \\
\frac{d c_1}{d u}&\equiv& \lim_{y \rightarrow \infty}\,\bar{\psi}_{,u}, \label{eq30}\\
\nonumber \\
\frac{d c_2}{d u}&\equiv& \frac{1}{2} \lim_{y \rightarrow \infty}\,\left(\mathrm{e}^{{2\bar{\psi}}/{y}}\bar{\omega}\right)_{,u}. \label{eq31}
\end{eqnarray}

\noindent To measure a deviation from the Bondi formula due to the numerical solution, we introduced the function $C(u)$ defined by 
\begin{eqnarray}
\mathcal{C}(u) &&= \left|\frac{1}{2\delta u}\left(M(u+\delta u)-M(u-\delta u)\right) + 
\left(\frac{d c_1}{d u}\right)^2+\left(\frac{d c_2}{d u}\right)^2 \right|. \label{eq32}
\end{eqnarray}

\noindent We approximate the derivative of the Bondi mass using the central difference scheme with $\delta u = h$, where $h$ is the step size of the numerical integration.
\begin{figure}[h]
	\begin{center}
		\includegraphics*[width=6.5cm,height=5.5cm]{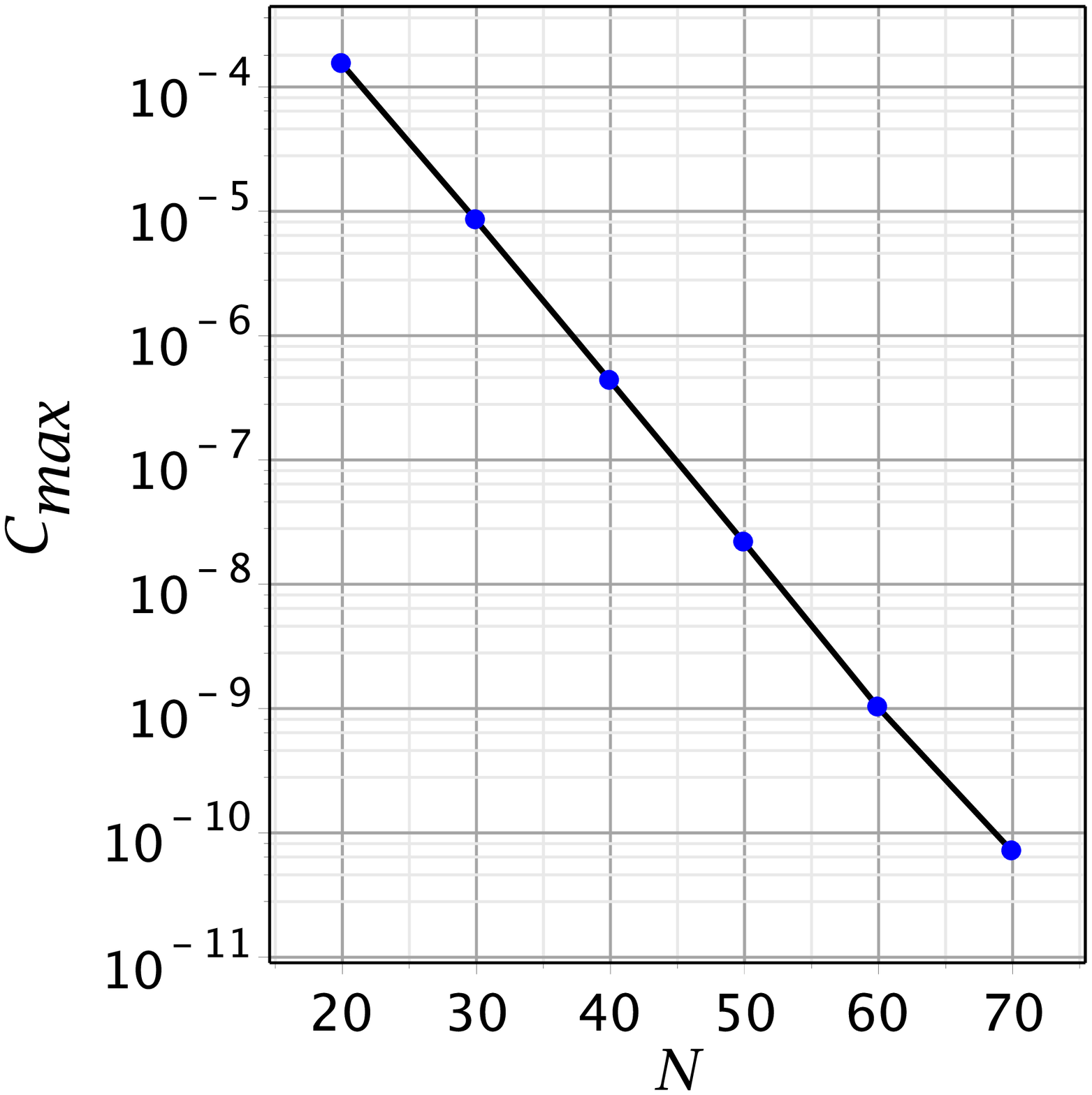}
		\includegraphics*[width=6.5cm,height=5.5cm]{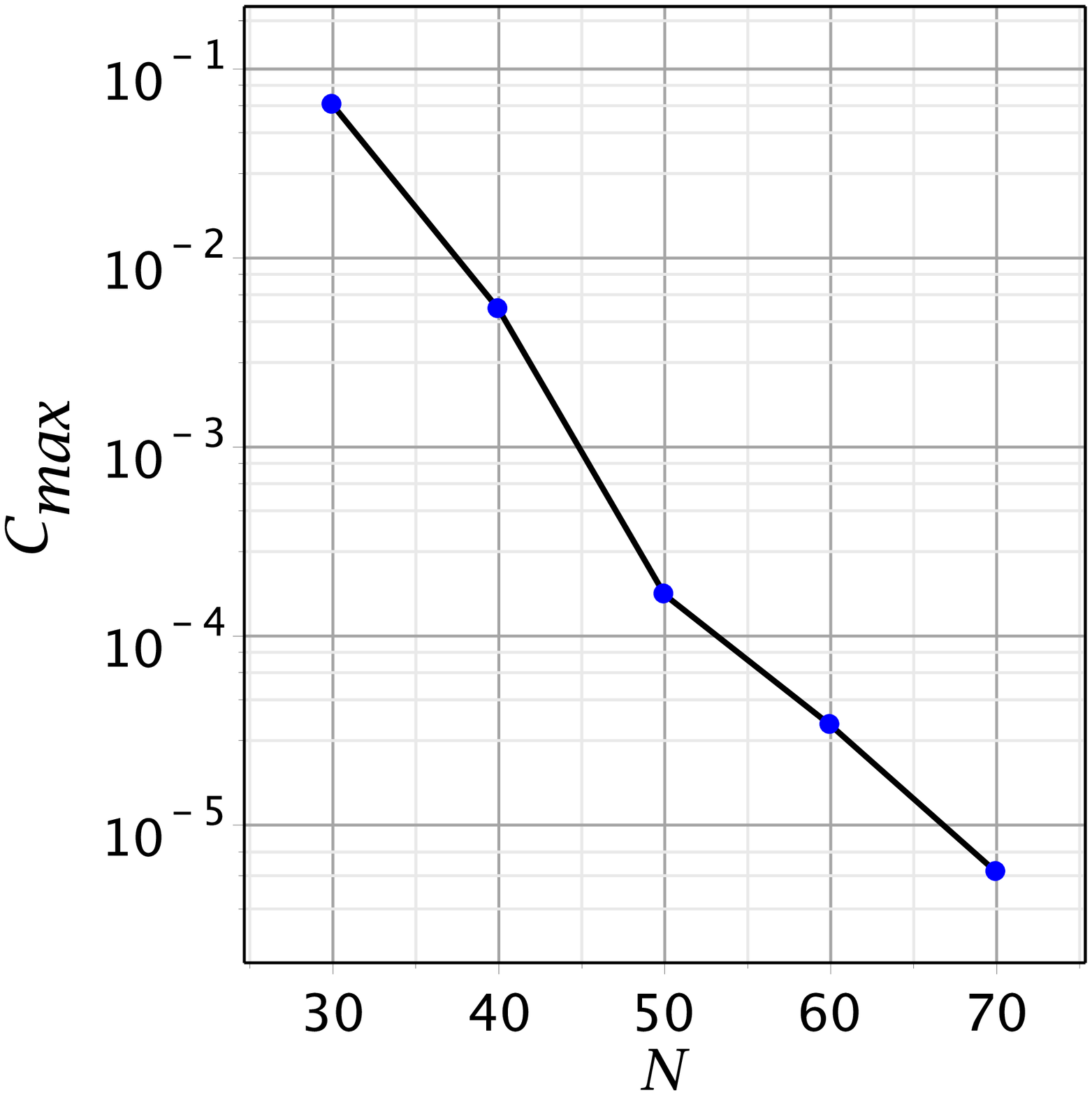}
	\end{center}
	{\renewcommand{\baselinestretch}{1}
		\vspace{-0.5cm}
		\caption{Exponential decay $C_{\mathrm{max}}$ (cf. Eq. (\ref{eq32})) for initial profiles of Fig. 2 (left panel) and Fig. 3 (right panel) in function of $N=N^{(A)}_\psi=N^{(A)}_\omega$, $A=1,2$}.} 
\end{figure}

The last numerical test consisted in determining the maximum value of $C(u)$ in function of $N=N^{(1)}+N^{(2)}$, where $N^{(A)}$, $A=1,2$ denote the truncation orders in each subdomain. We evolved the cylindrical waves with the initial configurations shown in Figs. 2 and 3 corresponding to the initial profiles of the Weber-Wheeler and the Xanthopoulos solutions for $u_0=-10$, respectively. For each truncation order in each domain, $N=N^{(A)}_\psi=N^{(A)}_\omega$, $A=1,2$, we collected the maximum deviation of the Bondi formula $C_{\mathrm{max}}$ given by Eq. (34). As depicted in Fig. 4, in both cases we obtain the exponential decay of $C_{\mathrm{max}}$.

\section{Templates of the gravitational waves}%

The peeling theorem plays a crucial role in characterizing the gravitational radiation emitted by an isolated source. After the works of Sachs \cite{sachs} and Newman and Penrose \cite{NP}, all the information containing in the Weyl tensor is expressed by five complex scalars known as the Weyl scalars. We denote these quantities by $\Psi_n$, $n=0,1,..,4$, obtained after a convenient projection of the Weyl tensor in a null complex tetrad basis. 

We can summarize the peeling theorem by the following behaviors of the Weyl scalars in a neighboorhood of the future null infinity $\mathcal{J}^{+}$ whose dominant term reads
\begin{eqnarray}
\Psi_n \simeq \frac{1}{r^{5-n}}, \label{eq33}
\end{eqnarray}

\noindent where $r$ is the affine parameter along the null rays. In particular, the Weyl scalar $\Psi_4$ falls off as $r^{-1}$ indicating that the gravitational field behaves like a plane wave asymptotically. Therefore, if distinct from zero, $\Psi_4$ provides a measure of the outgoing gravitational wave at the radiation zone, or at a large distance from the source. As a consequence, we can express the Peeling theorem \cite{sachs} in terms of the general asymptotic form of the Weyl tensor projected into the null tetrad basis (Weyl scalars) as
\begin{eqnarray}
C_{abcd} = \frac{N_{abcd}}{r} + \frac{III_{abcd}}{r^2} + \frac{II_{abcd}}{r^3} + \frac{I_{abcd}}{r^4} + ..., \nonumber \\
\label{eq34}
\end{eqnarray}

\noindent where the quantities $N_{abcd}, III_{abcd}, ...$ characterize the algebraic structure of the Riemann tensors of Petrov types $N, III, ...$, respectively. In particular, the spacetimes of type $N$ contain gravitational radiation with \begin{eqnarray}
\Psi_4 \simeq \frac{N_{abcd}}{r},  \label{eq35}
\end{eqnarray}  

\noindent implying that far from the source the curvature tensor has approximately the same algebraic structure of a Riemann tensor of a plane wave. Therefore, at the wave zone the templates of the outgoing gravitational radiation is described by $N_{abcd}$.

Stachel \cite{stachel} exhibited a version of the peeling theorem for the general cylindrical spacetimes, but the asymptotic expansion of the Riemann tensor is with inverse integer powers of $\rho^{\frac{1}{2}}$ since the spacetime is not asymptotically flat. It means that the Weyl scalars fall off as
\begin{eqnarray}
\Psi_n \simeq \frac{1}{y^{5-n}}. \label{eq36}
\end{eqnarray}

\noindent Hence, for cylindrical spacetimes the Weyl scalar $\Psi_4$ describes the outgoing gravitational radiation at the wave zone. By choosing a convenient null tetrad basis shown in the Appendix C, the following real and imaginary parts provides the template of the waves at the wave zone corresponding to the mode $+$ and $\times$, respectively
\begin{eqnarray}
(y\Psi_4)_\infty = \lim_{y \rightarrow \infty}\, \mathrm{e}^{-2\gamma} (2 \bar{\psi}_{,u} \gamma_{,u} - \bar{\psi}_{,uu})
+ i \lim_{y \rightarrow \infty}\, \frac{1}{2}\mathrm{e}^{-2\gamma} (-2 \bar{\omega}_{,u} \gamma_{,u} + \bar{\omega}_{,uu}). \label{eq37}
\end{eqnarray}

In this Section, we extract the numerical wave templates for some cases of interest after substituting the approximations given by Eqs. (\ref{eq15}) and (\ref{eq16}) ($A=2$) into Eq. (\ref{eq37}). The asymptotic expressions for $\gamma_{,u}(u,y)$ and $\gamma(u,y)$ are calculated from Eqs.  (\ref{eq7}), (\ref{eq8}) and (\ref{eq29}), respectively. The approximate $(y\Psi_4)_\infty$ is related to the modes $a^{(2)}_k, b^{(2)}_k$ whose evolution is dictated by the dynamical system given by Eq. (26) after setting the initial configuration. 
\begin{figure}[h]
	\begin{center}
		\includegraphics*[width=7.5cm,height=6.5cm]{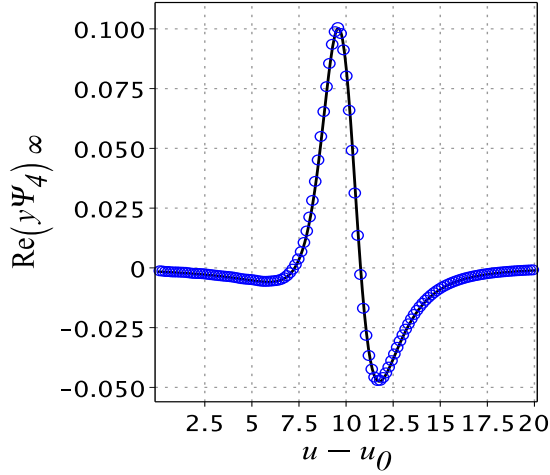}
	\end{center}
	\vspace{-0.5cm}
	\caption{Exact (continuous line) and numerical (circles) templates of the Weber-Wheeler solution. The initial configuration is the same of Fig. 2, and we have set $N_\psi^{(1)}=N_\psi^{(2)}=40$, and map paramter $L_0=2.0$. Here $u_0=-10$.	}
\end{figure}

The first case is the exact Weber-Wheeler solution describing a polarized gravitational wave that hits the symmetry axis to rebound back to infinity. The real part of Eq. (\ref{eq37}) determines the template at the radiation zone. Starting with the initial profile of Fig. 2, we present in Fig. 5 the exact (line) and the numerical (circles) plots of $\mathrm{Re}(y\Psi_4)_\infty$. The excellent agreement between the exact and numerical wave templates is another illustration of the accuracy of the domain decomposition algorithm (we chose truncation orders $N_\psi^{(1)}=N_\psi^{(2)}=40$). 

The second case corresponds to the evolution of polarized gravitational waves whose initial configurations are 

\begin{eqnarray}
\bar{\psi}_0(y) &=& A_0 y \mathrm{e}^{-{(y-y_1)^2}/{\sigma^2}}, \label{eq38a} \\
\nonumber \\
\bar{\psi}_0(y) &=& A_0 y^2 \left[1-\tanh \left(\frac{(y-y_2)^2}{\sigma^2}\right)\right], \label{eq38b}
\end{eqnarray}

\noindent where we fixed $\sigma=0.5$ and $y_1=y_2=3.0$. We obtained the templates of the outgoing gravitational radiation generated by the above initial data noticing that their structure is similar, but the details are distinct as shown in Fig. 6. In both cases, we have polarized waves $+$ that hits the axis and rebounce, and the additional structure in the patterns can be considered fingerprints of the particular initial data.
\begin{figure}[h]
\begin{center}
\includegraphics*[width=7.5cm,height=6.5cm]{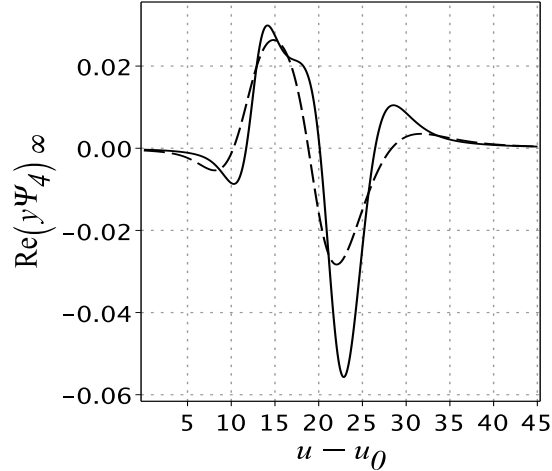}
\vspace{-0.5cm}
\end{center}
\caption{The wave templates for the Einstein-Rosen waves generated by the initial data families (40) (dashed line) and (41) (solid line). We have set $y_1=y_2=3.0$, $\sigma=0.5$, $y_0=L_0=2.5$ and $u_0=-10$. }
\end{figure}

In the next example, we obtain the templates resulting from the nonlinear interaction between the gravitational waves with polarization $+$ and $\times$. We choose the following initial data that contains a pure ingoing wave with polarization $\times$ (see Appendix D for details). It means that $\bar{\psi}(u_0,y)=0$, and 
\begin{equation}
\bar{\omega}(u_0,y)=\frac{B_0y}{1+y^4}\mathrm{e}^{-(y^2-\alpha_0^2)^2/\sigma^2} 
\end{equation}

\noindent where $B_0$ plays the role of the initial waves's amplitude, $\alpha_0$ and $\sigma$ are parameters.
\begin{figure*}[htb]
\begin{center}
         \includegraphics*[width=6.7cm,height=5.3cm]{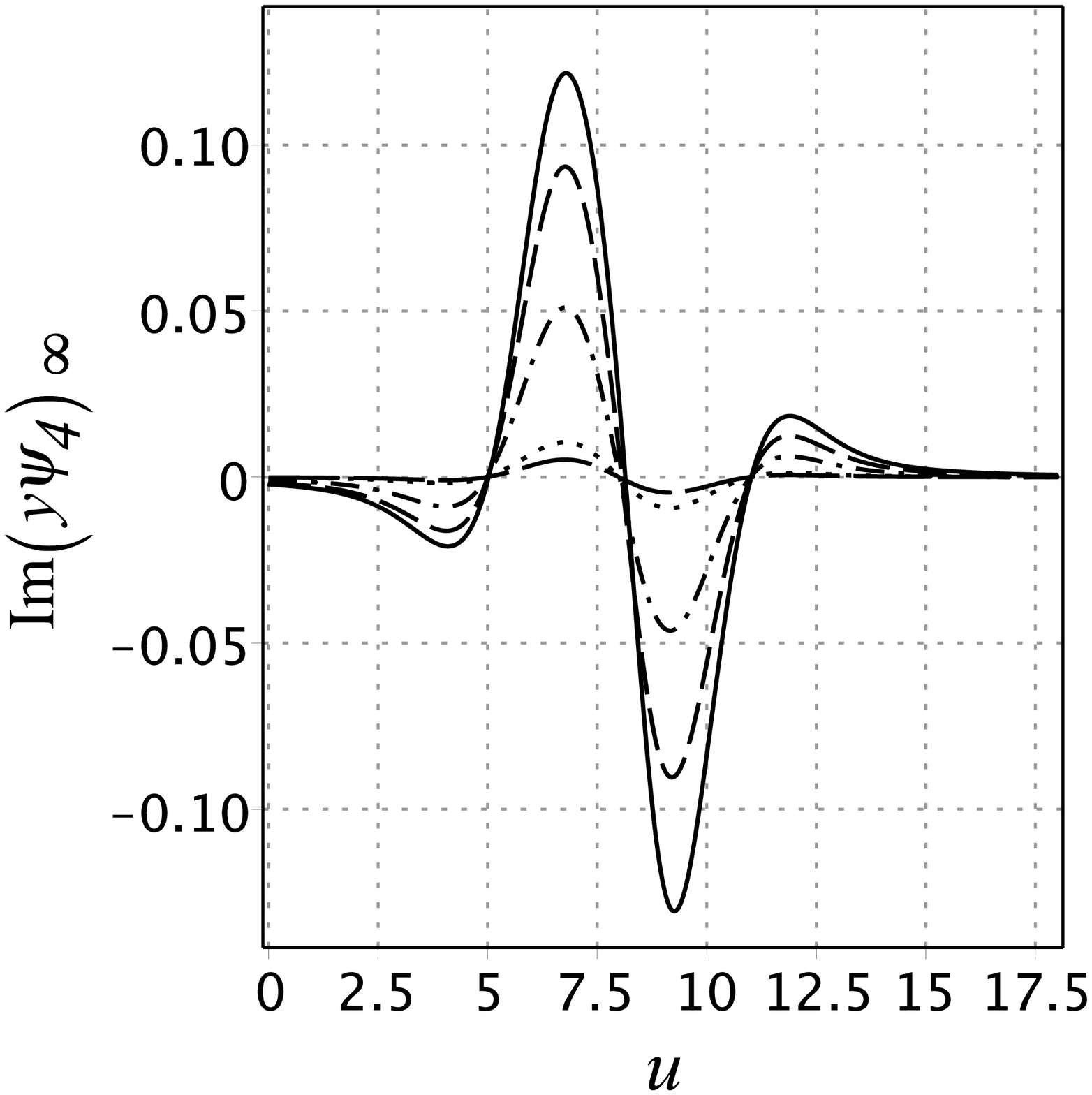}
         \includegraphics*[width=6.7cm,height=5.3cm]{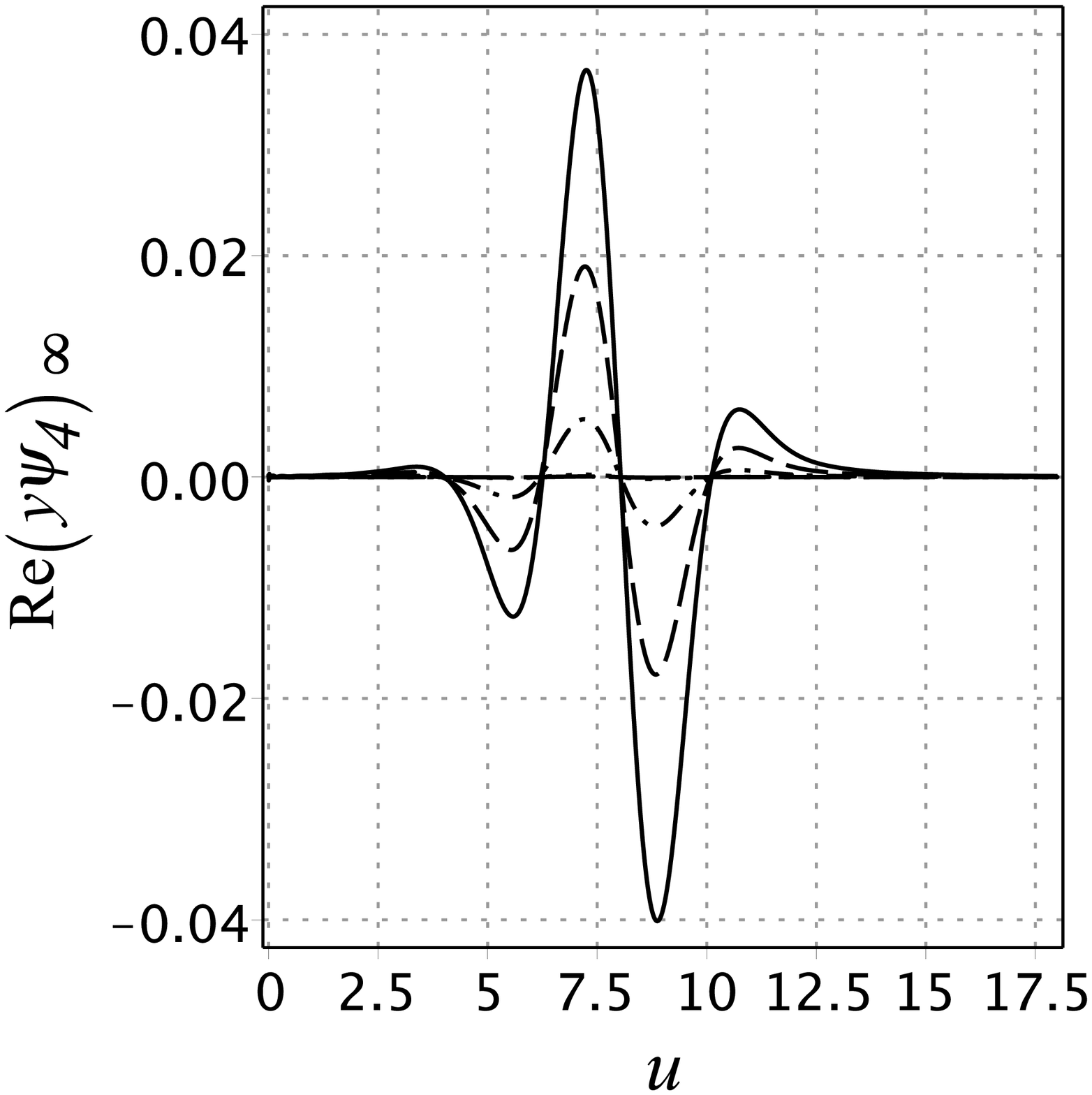}
\vspace{-0.5cm}
\center{(a)}	

\includegraphics*[width=6.7cm,height=5.3cm]{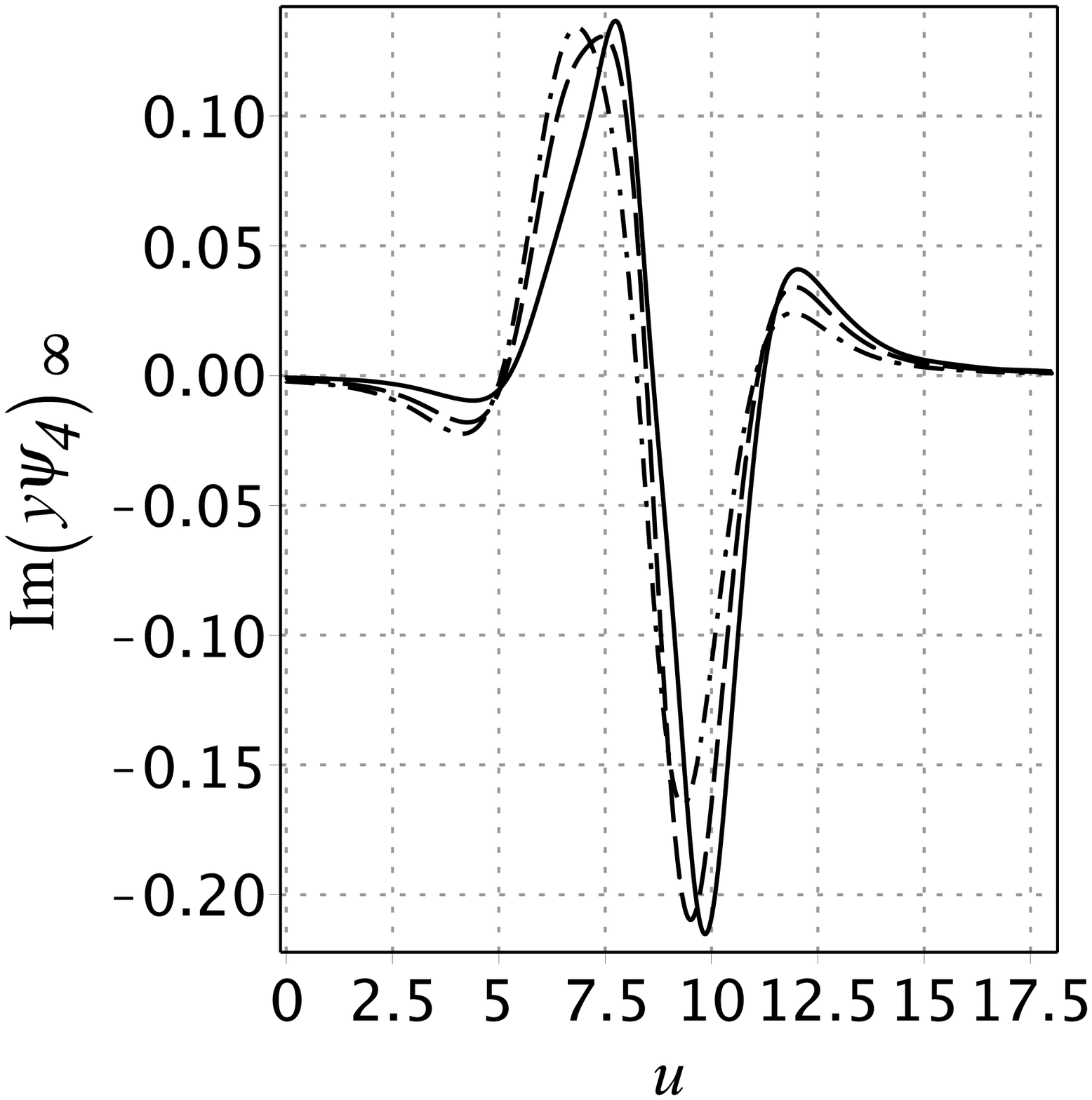}
\includegraphics*[width=6.7cm,height=5.3cm]{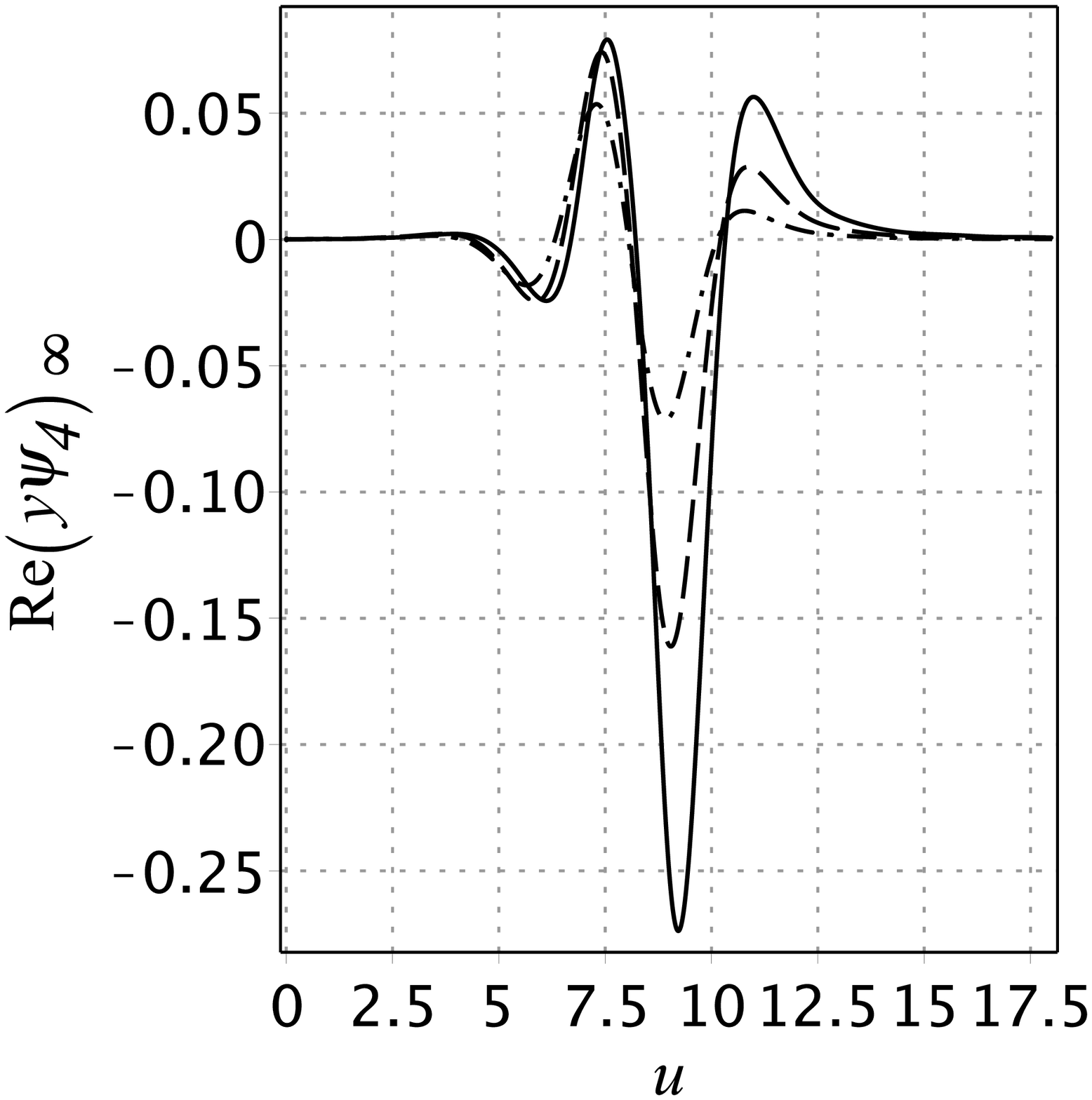}
\vspace{-0.5cm}
\center{(b)}

\includegraphics*[width=6.7cm,height=5.3cm]{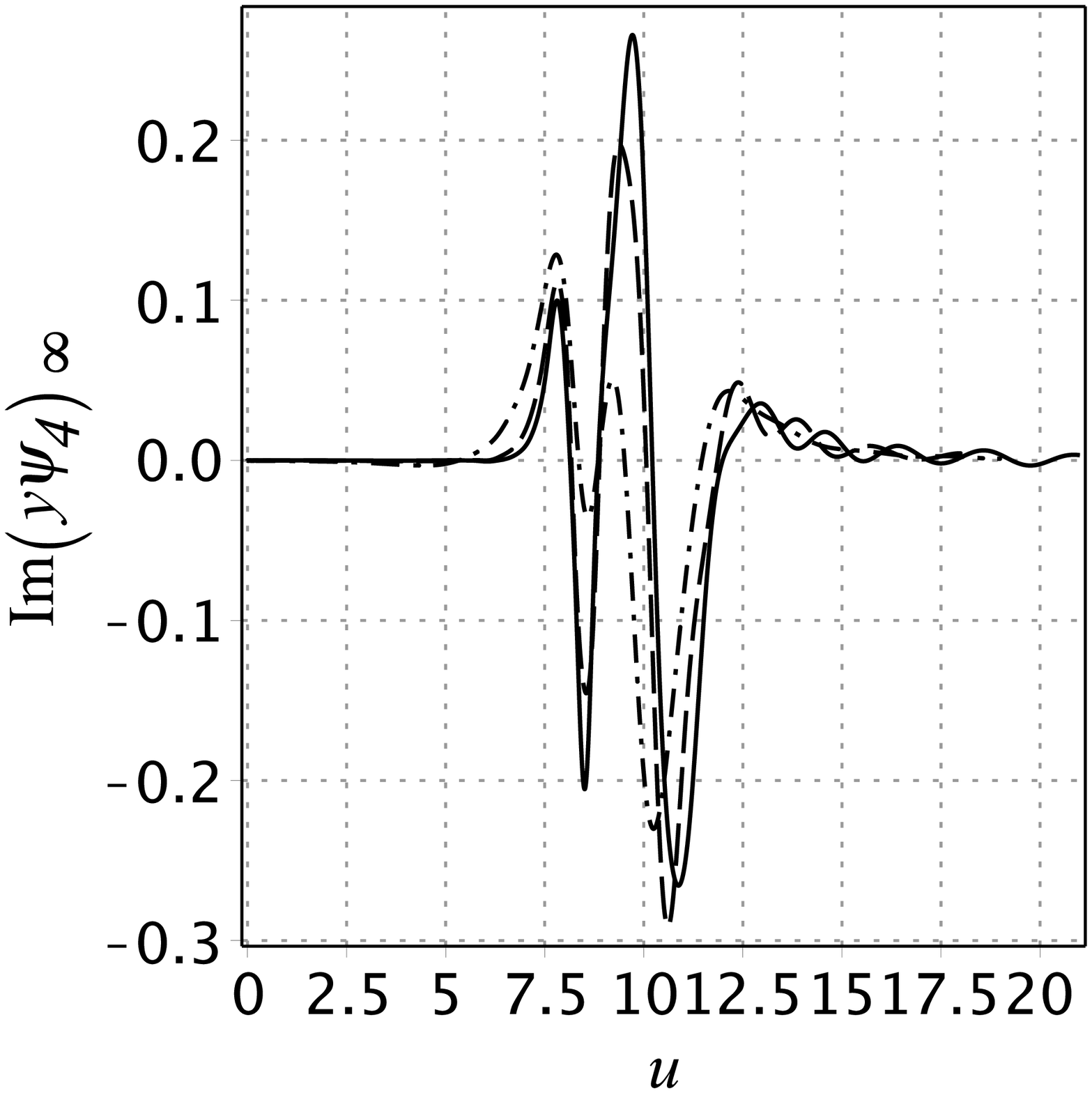}
\includegraphics*[width=6.7cm,height=5.3cm]{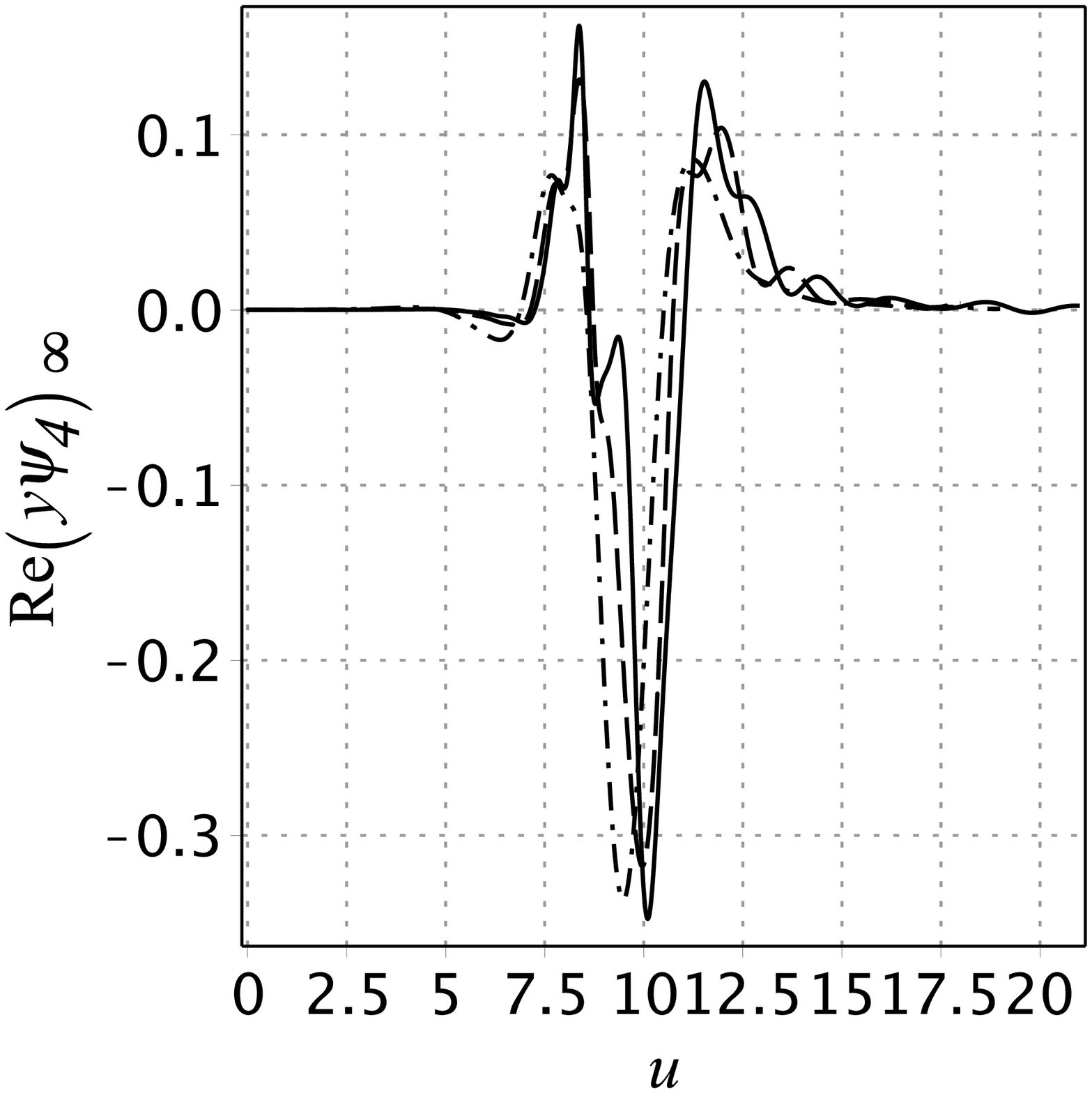}
\vspace{-0.5cm}
\center{(c)}
\end{center}
	{\renewcommand{\baselinestretch}{1}
		
\caption{(a) Wave templates for the modes $\times$ (left) and $+$ (right) for $B_0=0.05$ (long dash line), $B_0=0.1$ (dots) $B_0=0.5$ (dash dot line), $B_0=1.0$ (dash) and $B_0=1.5$ (solid line). (b) Wave templates for $B_0=2.0$ (dash dot), $B_0=3.0$ (dash) and $B_0=4.0$ (solide line). (c)  Wave templates for $B_0=5.0$ (dash dot), $B_0=6.0$ (dash) and $B_0=6.5$ (solid line).} 	}
\end{figure*}

When an initially ingoing gravitational wave with polarization $\times$ ($I_\times$) is directed towards the symmetry axes, the nonlinearity of the field equations enter into action producing the ingoing and outgoing wave modes $+$ ($I_+, O_+$), along with an outgoing wave mode $\times$ ($O_\times$). Therefore, the result is an unpolarized gravitational wave exhibiting templates described by $\mathrm{Re}(y \Psi_4)_\infty(u)$ and $\mathrm{Im}(y \Psi_4)_\infty(u)$. Moreover, Piran, Safier and Stark \cite{piran} showed an additional consequence of the interaction of both polarization modes characterized by the rotation between $I_+$ and $I_\times$, $O_+$ and $O_\times$, that is, the gravitational analog of the Faraday effect. It is, therefore, a valuable task to generate the wave templates in the present context. 

We evolved the spacetime starting with the initial data given by $\bar{\psi}(u_0,y)=0$ and Eq. (42) setting $\alpha_0=2$, $\sigma=1.0$ and increasing values of the amplitude $B_0$. We display in Fig. 7(a) the templates generated after choosing $B_0=0.05,\,0.1,\,0.5,\,1.0$ and $1.5$. The patterns associated with the wave modes $\times$ and $+$ oscillate with a phase shift as a consequence of the gravitational analog of the Faraday effect \cite{piran}. For $B_0=0.05$ and $0.1$ the average amplitudes of $\mathrm{Re}(y\Psi_4)_\infty$ are about $10^{-5}$ and $10^{-4}$, respectively, therefore both signals are indistinguishable in the plot. 

Increasing further $B_0$ the nonlinear coupling between the polarization wave modes $\times$ and $+$ turn to be more effective. With $B_0=2.0,\,3.0,\,4.0$ shown in Fig. 7(b), we notice that the average amplitude of the signal $\mathrm{Im}(y\Psi_4)_\infty$ almost does not change, while there is considerable growth in the average amplitude of $\mathrm{Re}(y\Psi_4)_\infty$. It is noteworthy the accentuated growth in the amplitude of the mode $+$ from about $10^{-5}$ for $B_0=0.05$ to $10^{-1}$ for $B_0=4.0$. In Fig. 7(c) the amplitudes are $B_0=5.0,\,6.0,\,6.5$. As noticed there is a change in the pattern structure of both polarization wave modes with more oscillations and the appearance of tails as a consequence of the strong reflection of the ingoing radiation and the rapid production of the ingoing and outgoing wave modes $+$. 

It is useful to illustrate the interaction of both polarization wave modes starting with an incoming wave $\times$ by presenting a sequence of 2-dimensional snapshots of $I_\times$ and $I_+$. The sequence showed in Fig. 8 corresponds to $B_0=2.0$ with $I_\times$ and $I_+$ represented by black (dash) and blue (solid) lines, respectively, and taken at $u=0,\,1,\,2,..,\,9$. Then, as the wave $\times$ propagates towards the symmetry axes, the wave $+$ emerges as a consequence of the nonlinear interaction of both wave modes. We remark that the growth of $I_+$ occurs out-of-phase in relation with $I_\times$. In this situation the gravitational Faraday effect takes place, meaning that the rotation of the polarization vector associated with the wave mode $+$ produces the phase shift imprinted in the templates at the wave zone.

\begin{figure}
\vspace{-1cm}
	\begin{center}
\includegraphics*[width=12cm,height=17cm]{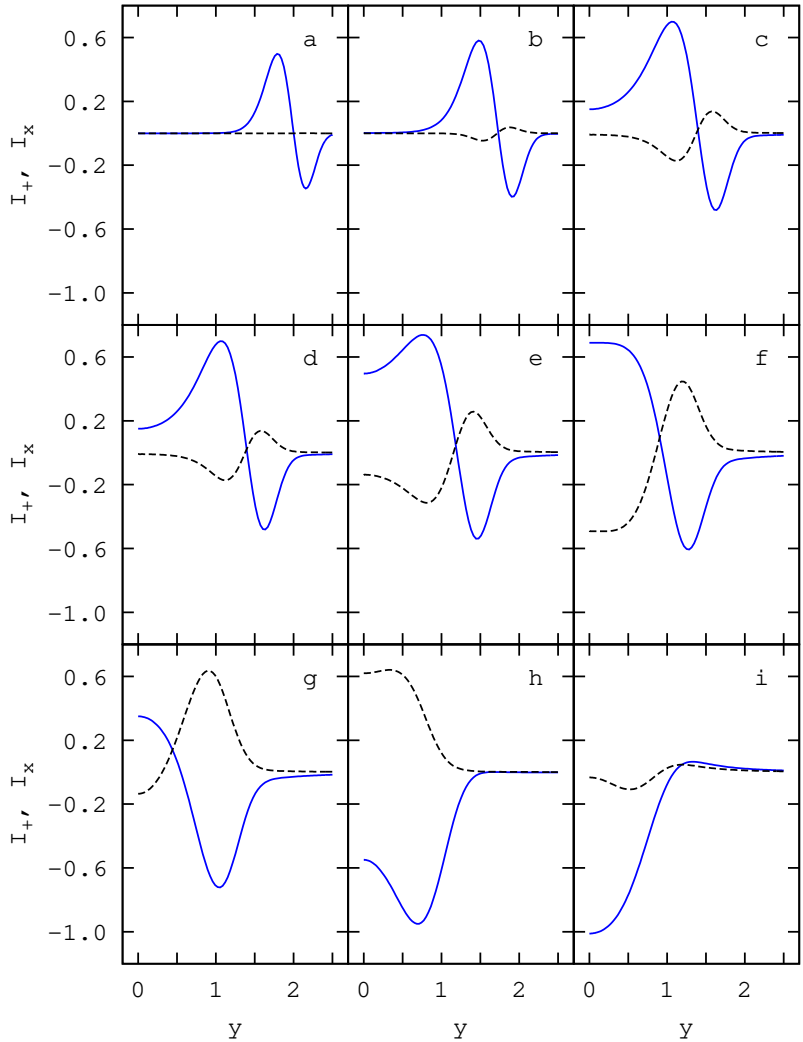}
	\end{center}
\vspace{-2.cm}
	\caption{Snapshots of $I_\times$ (solid lines) and $I_+$ (dash lines) for an incoming $\times$ wave with amplitude $B_0=2.0$ at $u=1.0,2.0,..,9.0$ from left to right and top to bottom.}
\end{figure}

\section{Final remarks}

We developed a version of the Galerkin-Collocation method with the technique of domain decomposition to evolve general cylindrical gravitational waves. The advantage of the domain decomposition over the single domain code becomes evident when the potentials $\bar{\psi}(u,y)$ and $\bar{\omega}(u,y)$ have high gradients in some regions of the spatial domain. 


We want to highlight two relevant aspects that differentiate from other codes. The first is the way we have introduced the computational domains schematically illustrated in Fig. 1 and described more precisely by the maps (\ref{eq13}) and (\ref{eq14}) that cover the whole spatial domain. In particular, the map (\ref{eq14}) is suitable for those functions that decay algebraically as $r \rightarrow \infty$ \cite{boyd}. The second aspect is the simple form of the basis functions that capture the behavior of the metric potentials near the origin and asymptotically. 

The numerical tests show the effectiveness of the domain decomposition algorithm. The first was to compare the approximate numerical initial data obtained in the single and double domains algorithms with the exact solutions of Weber-Wheeler and Xanthoupolos, where the initial data extracted from these solutions presented a region of steep variation. The second test was to evolve these data corresponding to polarized and non-polarized gravitational waves and keep track of the maximum deviation of the Bondi formula (33). By increasing the number of grid points in each domain, the error decays exponentially as expected. 

In the sequence, we exhibited the templates of cylindrical gravitational waves described by the Weyl scalar $\Psi_4$, with real and imaginary components corresponding to the wave modes of polarization $+$ and $\times$. The simplest case is of a polarized Einstein-Rosen wave (polarization $+$) that hits the axis and rebounce. The corresponding template has a basic structure not changed considerably by the initial data. The most interesting case is a pure ingoing wave with polarization $\times$ that, due to the nonlinearities entering into action, produces the ingoing and outgoing wave mode $+$. By increasing the initial amplitude of the ingoing wave mode $\times$ (parameter $B_0$) we noticed three main aspects. The wave templates $+$ and $\times$ oscillate with a phase shift possibly as a consequence of the gravitational analog of the Faraday effect. The second aspect is a rapid growth of the wave mode $+$ if compared with the growth in the amplitude of the template $\times$ when we increase $B_0$. And finally, both wave templates display a richer structure for a high amplitude initial incoming wave with polarization $\times$ which signalizes the strong field effect.

Finally, we point out that the present numerical scheme can be extended to study the nonlinear dynamics of gravitational waves in more general scenarios, such as provided by the Bondi problem. In particular, with the numerical domain decomposition code, we expect to explore the templates of the gravitational wave emission and, in the strong field regime, to follow the implosion of gravitational waves and the formation of black holes.

\ack{The authors acknowledge the financial support of the Brazilian agencies Conselho Nacional de Desenvolvimento Cient\'ifico e Tecnol\'{o}gico (CNPq) and Coordena\cao\ de Aperfei\c coamento de Pessoal de N\'ivel Superior (CAPES). H. P. O. and W. O. B. thank Funda\cao\ Carlos Chagas Filho de Amparo \`{a} Pesquisa do Estado do Rio de Janeiro (FAPERJ) for support within the Grants No. E-26/202.998/518 2016 and No. E-26/201.697/2018, Bolsas de Bancada de Projetos (BBP) and Bolsa de Pesquisador Visitante (BPV), respectively.}

\appendix

\section{Basis functions}  

First we define the auxiliary basis $\chi_k(y)$ as

{\small
\begin{eqnarray}
\chi_k(y) &=& \frac{1}{2}TL^{(1)}_{k+2}(y) + \frac{k+1}{2k+1} TL^{(1)}_{k+1}(y) + \frac{2k+3}{4(2k+1)} TL^{(1)}_{k}(y) 
\end{eqnarray}}

\noindent and the basis functions $\Phi^{(1)}_k(y)$ are

{\small
\begin{eqnarray}
\Phi^{(1)}_k(y) &=& \frac{1}{4} \chi_{k}(y) + \frac{(2k^2+5k+3)}{4 (2k^2+9k+9)} \chi_{k+1}(y).
\end{eqnarray}
}

\section{Exact solutions: the Weber-Wheeler and the Xanthopoulos solutions}

The particular Weber-Wheeler solution is given by \cite{weber_wheeler}

\begin{eqnarray}
&&\psi_{\mathrm{exact}}(u,y)=\nonumber \\
&&A_0 \left\{{\frac { \left\{a^2+y^4+(u+y^2)^2\,[2a^2-2y^4+(u+y^2)^2]\right\}^{1/2}+
a^2-u^2-2\,uy^2}{(a^2+y^4)^2+(u+y^2)^2\,[2a^2-2y^4+(u+y^2)^2]}}\right\}^{1/2}, \nonumber\\
\end{eqnarray}

\noindent where $A_0$ and $a$ are parameters identified as the amplitude and the width of the wave, respectively.


We can express the Xanthopoulos solution \cite{xanthopoulos} in terms of the null coordinates adopted here. Briefly, to this aim the necessary steps are the following. (i) First, we established the correspondence between the metric functions $q_2,\chi,\nu$ of Ref. \cite{xanthopoulos} and $\psi,\omega,\gamma$ of the line element (1). (ii) We obtained the relation connecting the ``prolate" coordinates $(\mu,\eta)$ (Eq. of \cite{xanthopoulos}) and the coordinates $(u,y)$. (iii) From the solution expressed in function of $(\mu,\eta)$ (Section IV of \cite{xanthopoulos}) we combine (i) and (ii) to obtain the expression for $\psi,\omega$ and $\gamma$ of the Xanthopoulos solution. These expressions are:

{\small
\begin{eqnarray}
\bar{\psi}_{\mathrm{exact}}(u,y) &=& \frac{y}{2} \log\left[\frac{p^2(\eta^2+\mu^2)+\mu^2+1}{(1-p\eta)^2+(1+p^2)\mu^2}\right], \\
\nonumber \\
\bar{\omega}_{\mathrm{exact}}(u,y) &=& \frac{2\sqrt{1+p^2}(\mu^2-1)(1-p\eta)}{p y\left[p^2(\eta^2+\mu^2)+\mu^2-1\right]},
\end{eqnarray}
}

\noindent where $p$ is a free parameter, $\mu$ and $\eta$ are functions of $(u,y)$

\begin{eqnarray}
\mu(u,y) &=&  \frac{\Delta}{\sqrt{2}},
\nonumber \\
\eta(u,y) &=& \frac{\sqrt{2}(u+y^2)}{\Delta},
\end{eqnarray}

\noindent and 

{\small
\begin{equation}
\Delta = \left[-u^2-2uy^2+1+\sqrt{\left(-u^2-2uy^2+1\right)^2+4(u+y^2)^2}\right]^{1/2}. 
\end{equation}
}

\noindent The second parameter, $\alpha$, appears in the expression of $\gamma(u,y)$ as

{\small
\begin{eqnarray}
\mathrm{e}^{2\gamma} = \frac{\alpha^2 \left[p^2(\eta^2+\mu^2)+\mu^2-1\right]}{\eta^2+\mu^2}.
\end{eqnarray}
}

\section{The null tetrad basis}

We can express components of the metric tensor with respect to a set of null tetrads as
\begin{eqnarray}
g_{\mu\nu} = -l_\mu k_\nu - k_\mu l_\nu + m_\mu \bar{m}_\nu + \bar{m}_\mu m_\nu,
\end{eqnarray}

\noindent where $l_\mu$, $k_\mu$ and $m_\mu$ are null vectors that satisfy the relations $l_u k^\mu = - m_u \bar{m}^\mu = -1$. Following Stachel \cite{stachel}, we have
\begin{eqnarray}
l_\mu &=& \mathrm{e}^{2(\gamma-\psi)}\,\delta^0_\mu, \\
k_\mu &=& \frac{1}{2}\,\delta^0_\mu+\delta^1_\mu = \left(\frac{1}{2},1,0,0\right), \\
m_\mu &=& \frac{1}{\sqrt{2}}\,\left(0,0,\mathrm{e}^\psi,\omega \mathrm{e}^\psi - i \rho \mathrm{e}^{-\psi}\right), \\
\bar{m}_\mu &=& \frac{1}{\sqrt{2}}\,\left(0,0,\mathrm{e}^\psi,\omega \mathrm{e}^\psi + i \rho \mathrm{e}^{-\psi}\right).
\end{eqnarray} 

\noindent The Newman-Penrose scalar $\Psi_4$ is given by
\begin{eqnarray}
\Psi_4 = R_{\mu\nu\alpha\beta} \bar{m}^\mu k^\nu \bar{m}^\alpha k^\beta,
\end{eqnarray}

\noindent and the corresponding real and imaginary parts are respectively
%
\begin{eqnarray}
(\Psi_4)^{\mathrm{real}} && = \mathrm{e}^{2(\psi-\gamma)} \left\lbrace \frac{1}{2y}\left(\frac{\bar{\psi}_{,u}}{y}\right)_{,y} - \frac{\bar{\psi}_{,uu}}{y} - \frac{1}{16y^2}\left(\frac{\bar{\psi}}{y}\right)_{,yy}+ \frac{1}{16y^3}\left(\frac{\bar{\psi}}{y}\right)_{,y} \nonumber \right.\\ 
& &  + \left[\frac{1}{2y}\left(\frac{\bar{\psi}}{y}\right)_{,y} - 2 \frac{\bar{\psi}_{,u}}{y} - \frac{1}{2 y^2}\right]\left(\frac{1}{4y} \gamma_{,y} - \gamma_{,u}\right)  - \frac{1}{2}\left(\frac{1}{2y} \left(\frac{\bar{\psi}}{y}\right)_{,y} - 2 \frac{\bar{\psi}_{,u}}{y}\right)^2 \nonumber \\
& &\left. + \frac{\mathrm{e}^{4 \psi}}{2 y^4}\left[\frac{1}{4 y} (y\bar{\omega})_{,y} - y \bar{\omega}_{,u}\right] \right\rbrace,
\end{eqnarray}

\begin{eqnarray}
(\Psi_4)^{\mathrm{im}} &&= \frac{1}{2y} \mathrm{e}^{2(\psi-\gamma)}\,\left\lbrace \left[\frac{1}{2 y^2}(y \bar{\omega})_{,y} - \bar{\omega}_{,u}\right]\left[\frac{3}{2y}\left(\frac{\bar{\psi}}{y}\right)_{,y} - \frac{6 \bar{\psi}_{,u}}{y} - \frac{\gamma_{,y}}{2 y} + 2 \gamma_{,u} - \frac{1}{2 y^2}\right]  \right. \nonumber \\
\nonumber \\
& &\left. - \frac{(y \bar{\omega}_{,u})_{,y}}{2 y^2} + \bar{\omega}_{,uu} + \frac{1}{16y^3} \left[(y\bar{\omega})_{,yy} - \frac{(y\bar{\omega})_{,y}}{y}\right] \right\rbrace.
\end{eqnarray}

\section{The initial data of ingoing $\times$ waves}

We follow Piran et al. \cite{piran} the quantities $I_+,\,O_+,\,I_\times$ and $O_\times$ denote the amplitude of the ingoing and outgoing waves in the modes $+$ and $\times$, respectively. We can rewrite these quantities in null coordinates and using the redefined potentials $\bar{\psi}$ and $\bar{\omega}$ as:
{\small
\begin{eqnarray}
I_+=2(\psi_{,t}+\psi_{,\rho}) = \frac{1}{y}\left(\frac{\bar{\psi}}{y}\right)_{,y},\;\;O_+=2(\psi_{,t}-\psi_{,\rho})=2\left[2\left(\frac{\bar{\psi}}{y}\right)_{,u}-\frac{1}{2y}\left(\frac{\bar{\psi}}{y}\right)_{,y}\right], \nonumber \\
\\
I_\times=\frac{\mathrm{e}^{2 \psi}}{\rho}(\omega_{,t}+\omega_{,\rho}) =\frac{\mathrm{e}^{ \frac{2\bar{\psi}}{y}}}{2 y^3}\left(y\bar{\omega}\right)_{,y},\;O_\times=\frac{\mathrm{e}^{2 \psi}}{\rho}(\omega_{,t}-\omega_{,\rho})=\frac{\mathrm{e}^{ \frac{2\bar{\psi}}{y}}}{2 y^2}\left[2(y\bar{\omega})_{,u}-\frac{(y\bar{\omega})_{,y}}{2y}\right]. \nonumber 
\end{eqnarray}
}

We derived the initial data corresponding to pure ingoing $\times$ polarized waves by setting $I_+=O_+=0$ initially, implying that $\bar{\psi}(u_0,y)=0$, and

\begin{equation}
\frac{\mathrm{e}^{ \frac{2\bar{\psi}}{y}}}{2 y^2}\left[2(y\bar{\omega})_{,u}-\frac{(y\bar{\omega})_{,y}}{2y}\right] = 0,
\end{equation}

\noindent at $u=u_0$. The general solution of the above equation is 

\begin{equation}
\bar{\omega}(u,y)=\frac{B_0}{y} F\left(y^2+\frac{1}{2}(u-u_0)\right), 
\end{equation}

\noindent where $F$ is an arbitrary function that is consistent with the boundary conditions for $\bar{\omega}$ given by Eqs. (10) and (11). A convenient choice for $F$ at $u=u_0$ is $F(y^2)=y^4/(1+y^4) \mathrm{e}^{-(y^2-\alpha_0^2)^2/\sigma^2}$, where $\alpha_0$ and $\sigma$ are arbitrary parameters. Thus, the initial data describing incoming $\times$ gravitational waves initially is

\begin{eqnarray}
\bar{\psi}(u_0,y)=0 , \\
\nonumber \\
\bar{\omega}(u_0,y)=\frac{B_0 y}{1+y^4}\mathrm{e}^{-(y^2-\alpha_0^2)^2/\sigma^2}.
\end{eqnarray}

\end{document}